\documentclass[twocolumn]{aa}
\usepackage{graphicx,epsfig}

\begin{document}

   \title{Biased total mass of cool core galaxy clusters by Sunyaev-Zel'dovich effect measurements}
   \subtitle{}

   \author{A. Conte \inst{1}, M. De Petris\inst{1}, B. Comis\inst{1}, L. Lamagna\inst{1}, and S. De Gregori\inst{1}
          }

   \offprints{andrea.conte@roma1.infn.it}

   \institute{\inst{1}Sapienza University, Department of Physics,
              P.le A. Moro, 2 Rome, Italy\\
             \email{andrea.conte@roma1.infn.it}
             }

   \date{Accepted xx. Received yy; in original form zz}

  \abstract{The Sunyaev Zel'dovich (SZ) effect from galaxy clusters is one of the most powerful
cosmological tools for investigating the large-scale Universe. The
big advantage of the SZ effect is its redshift independence, which
is not the case for visible and X-ray observations. It allows us to
directly estimate the cluster's total mass from the integrated
comptonization parameter $Y$, even for distant clusters. However,
not having a full knowing intra-cluster medium (ICM) physics can
affect the results. By taking self-similar temperature and density
profiles of the ICM into account, we studied how different ICM
morphologies can affect the cluster total mass estimation. With the
help of the high percentage of cool core (CC) clusters, as observed
so far, the present analysis focuses on studying this class of
objects. A sample of eight nearby ($0.1<z<0.5$) and high-mass
($M>10^{14}$ $M_{\odot}$) clusters observed by Chandra was
considered. We simulated SZ observations of these clusters through
X-ray derived information and analyzed the mock SZ data again with
the simplistic assumption of an isothermal \textit{beta}-model
profile for the ICM. The bias on the recovered cluster total mass
using different sets of assumptions is estimated to be 50\% higher
in the case of hydrostatic equilibrium. Possible contributions to
the total bias due to the line-of-sight integration and the
considered ICM template are taken into account. The large biases on
total mass recovery firmly support, if still necessary, cluster
modeling based on more sophisticated universal profiles as derived
by X-ray observations of local objects and hydrodynamical
simulations.
  }
\titlerunning{Biased total mass of CC galaxy clusters by SZE measurements}
\authorrunning{Conte et al.}
   \maketitle
%

\section{Introduction}

\begin{table*}
\begin{minipage}[htbp]{\textwidth}
\caption{Parameters of the electron number density
profiles.}\label{table_neCC} \centering
\renewcommand{\footnoterule}{}  
  \begin{tabular}{lcccccc}
    \hline
    Cluster &  $n_{e0}$ & $\beta$ & $\theta_{c1}$ & $\theta_{c2}$ & $f$ \\
      & (10$^{-2}$cm$^{-3}$) &  & (arcsec) & (arcsec) &  \\
    \hline
    \hline
    A1413 & $3.89\pm0.54$ & $0.535\pm0.016$ & $6.7\pm1.4$ & $40.1\pm4.1$ & $0.760\pm0.020$ \\
    A1689  & $4.15\pm0.31$ & $0.871\pm0.040$ & $21.6\pm1.0$ & $104.5\pm5.3$ & $0.870\pm0.010$ \\
    A1835  & $11.3\pm0.4$ & $0.802\pm0.015$ & $9.3\pm0.2$ & $63.8\pm1.6$ & $0.940\pm0.001$ \\
    A2204  & $20.4\pm1.1$ & $0.716\pm0.028$ & $7.5\pm0.3$ & $67.6\pm1.9$ & $0.959\pm0.004$ \\
    A2261  & $4.07\pm0.59$ & $0.631\pm0.024$ & $10.2\pm1.8$ & $39.1\pm5.9$ & $0.760\pm0.050$ \\
    MS1358.4+6245  & $9.63\pm0.79$ & $0.676\pm0.017$ & $3.3\pm0.2$ & $37.0\pm1.8$ & $0.934\pm0.003$ \\
    RXJ1347.5-1145  & $28.5\pm1.4$ & $0.632\pm0.009$ & $4.0\pm0.2$ & $23.3\pm1.6$ & $0.942\pm0.004$ \\
    ZW3146  & $16.9\pm0.3$ & $0.669\pm0.005$ & $4.4\pm0.1$ & $25.8\pm0.6$ & $0.882\pm0.004$ \\
    \hline
  \end{tabular}
\end{minipage}
\end{table*}

Clusters of galaxies are the largest gravitationally-bound objects
arising thus far from the process of hierarchical structure
formation (\cite{Voit05}). As the most recent and most massive
objects of the Universe, clusters are excellent probes for studying
its formation and evolution. The observed state of gas within a
cluster is determined by a combination of shock heating during
accretion, radiative cooling, and thermal feedback produced by the
cooling itself, so the density and temperature of the ICM represent
the full thermal history of clusters' formation. To better
understand the physics of ICM, it is necessary to have sufficient
knowledge of the gas density and temperature distributions. Though
clusters are the ideal target objects for X-ray observations of the
hot ICM, millimeter and sub-millimeter measurements provide
independent and complementary tools for studying the same ICM by
exploiting the Sunyaev Zel'dovich (SZ) effect (\cite{Sunyaev72}).

The SZ effect is the Comptonization of the cosmic microwave
background (CMB) photons, coming from the last scattering surface,
by the hot electrons population of the ICM. The photon energy
variation, which is caused by the scattering process, can be
expressed as CMB temperature variations

\begin{equation}\label{eq_SZ}
\Delta T_{SZ}=yT_{CMB}f(x)(1+\delta_n(x,\theta_e))+\Delta T_{kin}
\end{equation}
where

\begin{equation}\label{eq_y}
y=\int\theta_ed\tau_e=\int\left(\frac{k_BT_e}{m_ec^2}\right)\sigma_Tn_edl\propto\int
P_edl
\end{equation}
represents the comptonization parameter, $x=(h\nu)/(k_BT_{CMB})$ the
dimensionless frequency, $h$ and $k_B$ are respectively the Planck
and Boltzmann constants, $T_{CMB}$, $m_e$ and $\sigma_T$, the CMB
temperature at $z=0$, the electron mass at rest and the Thomson
cross section, $\theta_e$ represents the dimensionless thermal
energy of the ICM, $\tau_e$ is the electron optical depth. The
parameters $n_e$, $T_e$, and $P_e$ are the electron number density,
temperature, and pressure of the ICM,
$\delta_n(x,\theta_e)=f_n(x)\theta_e^n/f(x)$ is the relativistic
correction term that accounts for the thermal energy of the
electrons involved in the scattering processes, where
$f(x)=x[(e^x+1)/(e^x-1)-4]$ is a dimensionless quantity that
describes the spectral signature of the effect, and the subscript
$n$ indicates the maximum order of the relativistic correction
($n=4$ in this work, \cite{Nozawa95}). The last term of Eq.
\ref{eq_SZ} is the kinematic component of the SZ effect, which
contains the contribution from the bulk motion of the electron
population with respect to the last scattering surface reference
frame. For the purpose of this paper, this term is omitted, assuming
that it is disentangled from the thermal component by
multi-frequency observations, together with the signal from the
primary CMB emission.

The SZ effect is redshift independent and, for this reason, it is
possible to detect distant clusters without any existing X-ray or
optical observations. This is the case of the ongoing ground-based
experiments such as SPT (\cite{Ruhl04}), ACT (\cite{Kosowsky03}),
and the all sky survey like Planck (\cite{Planck11a}) or the
upgraded MITO (\cite{DePetris07}) and OLIMPO (\cite{Masi08}) with
new spectroscopic capabilities and the proposed 30-m diameter C-CAT
(\cite{Sebring06}). However, some assumptions on cluster physics
still have to be made in order to directly extract cluster
observables.

Estimates of cluster's total mass can be derived by SZ observations
when X-ray or lensing measurements are available or by empirically
calibrated scaling relations linking the SZ flux to the total mass
(e.g. \cite{Vikhlinin09, Arnaud10, Planck11b, Comis11}). Total mass
can also be determined by SZ observations alone when applying
thermal energy constraints (\cite{Mroczkowski11}).

To accurately reproduce the gas inside the cluster, an ICM universal
model is mandatory (e.g. \cite{Nagai07, Arnaud10}). In this paper we
confirm that the simple isothermal \textit{beta}-model is clearly an
inappropriate cluster representation for total mass recovery by SZ
observations, particularly in the presence of relaxed cool core (CC)
clusters. These objects show a well studied peaked density profile
with a temperature decrement in the core region (\cite{Jones84}). In
the local universe this class of clusters is observationally a
significant percentage of the total cluster population
(\cite{Eckert11}). Even if the X-ray estimated CC fraction is biased
by selection effects in flux-limited samples, recently a 35\% of
clusters have picked up in the SZ high signal-to-noise ratio Planck
early cluster data-set (\cite{Planck11c}) are CC clusters. Large
scatter in mass estimates of CC clusters has been highlighted
previously using numerical simulations by Hallman et al. (2006) and
Hallman et al. (2007).

We investigate the bias on the mass in a limited sample of eight
nearby ($0.1<z<0.5$) and high-mass ($M>10^{14}$ $M_{\odot}$) CC
clusters observed by Chandra. The SZ maps of these clusters, which
are expressed in thermodynamic temperature units and convolved with
several instrumental beams, are dealt with by applying the
isothermal \textit{beta}-model. The total mass is derived in three
different ways: by assuming hydrostatic equilibrium and a fixed gas
fraction and by applying a self similar scaling relation. To focus
only on the consequences of the employed ICM model, in our analysis
we neglect all the contaminants present in the sky by assuming in
this way the best situation to recover cluster total mass.

In Sect. \ref{sec_ne_Te}, we discuss the electron number density
radial profile and follow self-similar studies to characterize a
universal electron temperature radial profile of a limited sample of
eight CC clusters observed by Chandra. In Sect. \ref{sec_MCMC} we
generate maps of the SZ effect in thermodynamic temperature. In
Sect. \ref{sec_total_mass} we evaluate cluster total mass under
different sets of assumptions. The bias on the recovered mass is
described in Sect. \ref{sec_MB}, which discusses the main
contributions. Conclusions are summarized in Sect.
\ref{sec_conclusions}.

\section{Electron number density and temperature
profiles}\label{sec_ne_Te}

\begin{table*}
\begin{minipage}[hbpt]{\textwidth}
\caption{CC galaxy clusters properties used in the analysis to
generate a universal $T_e$ profile for this class of clusters.}
\label{table_te} \centering
\renewcommand{\footnoterule}{}  
  \begin{tabular}{l c c c c c c}
    \hline
    Cluster & z & $D_A$ & $r_{500}$\footnote{\cite{Morandi07}} & $\theta_{500}$ & $T_X$\footnote{Temperature scale  calculated by a weighted mean of Bonamente et al. (2006) data.}& $T_{e0}$\footnote{Electron temperature of the cluster obtained by fitting the isothermal \textit{beta}-model to X-ray data (\cite{Bonamente06}).}\\
    name &  & (Gpc) & (kpc) & (arcsec) & (keV)& (keV)\\
    \hline
    \hline
    $A1413$ & 0.142 & 0.52 & $1195\pm232$ & $321\pm62$ & $6.6\pm0.6$& $7.3\pm0.2$\\
    $A1689$ & 0.183 & 0.63 & $1402\pm260$ & $377\pm70$ & $8.7\pm0.9$& $10.0\pm0.3$\\
    $A1835$ & 0.252 & 0.81 & $1439\pm414$ & $387\pm111$ & $10.5\pm1.0$& $8.4\pm0.2$\\
    $A2204$ & 0.152 & 0.55 & $1796\pm320$ & $483\pm86$ & $11.3\pm1.8$& $6.5\pm0.2$\\
    $A2261$ & 0.224 & 0.74 & $1201\pm168$ & $323\pm45$ & $7.0\pm1.0$& $7.2\pm0.4$\\
    $MS1358.4+6245$ & 0.327 & 0.97 & $1633\pm885$ & $439\pm238$ & $8.4\pm1.1$& $8.3\pm0.6$\\
    $RXJ1347.5-1145$ & 0.451 & 1.19 & $1734\pm170$ & $466\pm46$ & $14.8\pm1.5$& $13.5\pm0.5$\\
    $ZW3146$ & 0.291 & 0.90 & $1804\pm344$ & $485\pm92$ & $8.7\pm0.4$& $6.6\pm0.1$\\
    \hline
  \end{tabular}
\end{minipage}
\end{table*}

A general parametric model for the cluster atmosphere must be
defined to forecast the shape of cluster SZ signals in matched
filter techniques for detecting clusters in blind surveys. ACT has
detected new clusters assuming a two-dimensional Gaussian profile as
filter (\cite{Sehgal10}), while SPT has detected a projected
spherical \textit{beta} profile (\cite{Vanderlinde10}) and Planck a
universal pressure profile (\cite{Melin11}).

The approach for determining the total mass cluster can be
different. High-quality X-ray data allows an accurate modeling of
cluster morphology, but in the case of low angular resolution and/or
low signal-to-noise ratio a simple isothermal \textit{beta}-model is
still applied (e.g. \cite{Marriage10, Sayers11}).

In this work we analyze this model (\cite{Cavaliere78}), which is
based on the very general assumption that the electron temperature
$T_e$ is constant along the whole considered cluster radial
extension and that the electron number density follows a spherical
distribution as

\begin{equation}\label{eq_beta}
n_{e,ISO}(r)=n_{e0}{\left(1+\frac{r^2}{r_c^2}\right)}^{-\frac{3}{2}\beta},
\end{equation}
where $n_{e0}$ is the central electron number density, $r_c$ the
core radius, and $\beta$ the power law index. The subscript ISO
indicates, hereafter, the isothermal \textit{beta}-model. The proved
inadequacy of this model is compensated for by the advantage of
extracting a simple analytic expression for the $y$ parameter along
the off-axis angular separation, $\theta$,

\begin{equation}\label{eq_ybeta}
y_{ISO}(\theta)=y_0{\left(1+\frac{\theta^2}{\theta_c^2}\right)}^{\frac{1}{2}-\frac{3}{2}\beta}
\end{equation}
where

\begin{equation}\label{eq_y0}
y_0=n_{e0}\frac{k_BT_e}{m_ec^2}\sigma_T
r_c\sqrt{\pi}\frac{\Gamma\left(\frac{3}{2}\beta-\frac{1}{2}\right)}{\Gamma\left(\frac{3}{2}\beta\right)},
\end{equation}
and $\theta_c=r_c/D_A$, with $D_A$ the angular diameter distance,
which has been calculated for each cluster by using

\begin{equation}
D_A=\frac{c}{H_0(1+z)}\int^z_0\frac{dz'}{E(z')},
\end{equation}
where $H_0$ is the Hubble constant and
$E(z)={\left[\Omega_{M}(1+z)^3+\Omega_{\Lambda}\right]}^{1/2}$. We
adopt a $\Lambda CDM$ cosmology with $H_0=70$ km/s/Mpc,
$\Omega_M=0.3$, and $\Omega_{\Lambda}=0.7$.

It is easy to find clusters that are not relaxed or that display
structures that are very difficult to describe with this model.
Therefore, it is realistic to assume that many newly discovered
clusters in blind SZ surveys exhibit such significant deviations as
well. To explore a particular ICM gas morphology, we focus on CC
clusters. We started analyzing a small sample of eight objects
extracted from the Chandra dataset investigated in Bonamente et al.
(2006).

The study of a central region, commonly known as the core region,
has been challenged by high-resolution numerical simulations
(\cite{Navarro95, Borgani04, Kay04, Nagai07, Henning09}). These
works lead to an agreement on whether there is a cooling core in the
very central denser gas region ($r<0.1r_{500}$) of some clusters, as
well as a slower decline in the temperature at large radii
($r>0.2r_{500}$). As usual we refer to $r_{500}$ as the radius of
the cluster that defines a volume with mean density $500$ times the
critical density $\rho_{crit}$ at cluster redshift. The choice of
$r_{500}$ is motivated by simulation results from Evrard et al.
(1996) showing the gas within this radius relaxed and in hydrostatic
equilibrium.

Moreover, many observational studies in X-rays have shown that the
X-ray surface brightness, hence the underlying density, cannot be
represented correctly by a \textit{beta} profile. A second component
should be added or a peaked central part introduced in order to
properly fit the observation. The observed deprojected density
profiles are peaked for CC systems and flatter for morphologically
disturbed clusters.

The $n_e$ cluster profile can be described by a double
\textit{beta}-model profile

\begin{eqnarray}\label{eq_2beta}
n_{e,CC}(r)&=&n_{e0}\left[f{\left(1+\frac{r^2}{r_{c1}^2}\right)}^{-\frac{3}{2}\beta}+(1-f)
{\left(1+\frac{r^2}{r_{c2}^2}\right)}^{-\frac{3}{2}\beta}\right],\nonumber\\
\end{eqnarray}
where the parameters' data have been taken from Bonamente et al.
(2006) and adapted to this work to have a symmetric standard
deviation (D'Agostini, 2003).

This distribution is a generalization of a double
\textit{beta}-model profile of the electron number density,
developed by La Roque et al. (2005), but instead using the same
$\beta$ parameter for both the central region and the outskirts, as
in Bonamente et al. (2006). The $r_{c1}=\theta_{c1}/D_A$ and
$r_{c2}=\theta_{c2}/D_A$ are the core radii of the inner and outer
distributions, and $f$ is a parameter defined between 0 and 1 that
represents how the core region dominates the outer region. These
parameters, together with $n_{e0}$, are taken from Bonamente et al.
(2006) and summarized in Table \ref{table_neCC} for the selected
clusters.

\begin{figure}
\centering
\includegraphics[width=\columnwidth]{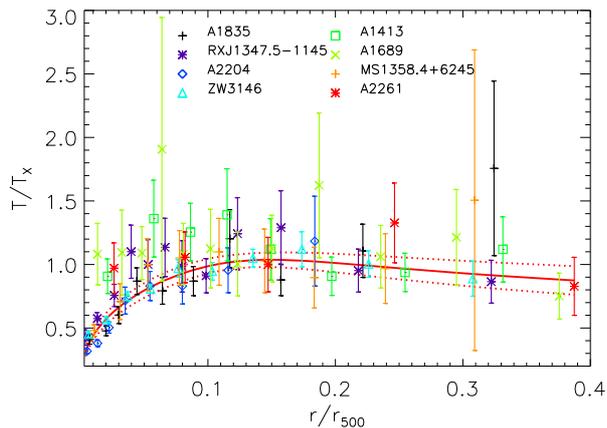}\\
  \caption{\itshape Radial electron temperature profiles of a Chandra-selected sample of CC galaxy clusters (\cite{Bonamente06}) and the best
  fit (solid line) of the electron temperature profile with the 1$\sigma$ error
  (dotted lines) proposed in this work. Temperatures and radii
  are expressed in terms of $T_X$ and $r_{500}$, respectively, which
  are used as scale quantities throughout this work.}\label{fig_fitTe}
\end{figure}

We describe the temperature profile as

\begin{eqnarray}\label{eq_Te_general}
\frac{T_{e,CC}(r)}{T_X}=\left\{
\begin{array}{rl}
A_1{\left(\frac{r}{r_{500}}\right)}^{m_1} & \mbox{  for } \frac{r}{r_{500}}<0.1 \\
A_2{\left(\frac{r}{r_{500}}\right)}^{m_2} & \mbox{  for }
\frac{r}{r_{500}}>0.2
\end{array}
\right.
\end{eqnarray}
where $A_{1,2}$ are determined by fixing the position at which the
two power laws, described by the $m_1$ and $m_2$ parameters,
intersect each other, as explained in Appendix \ref{sec_appendix}.

The search for a universal temperature profile in the cluster halo
region has been a target of several works based on observations
(e.g. \cite{Markevitch98, DeGrandi02, Zhang04, Vikhlinin05,
Vikhlinin06, Sanderson06, Zhang07, Pratt07, Zhang08}) and
hydrodynamical simulations (e.g., \cite{Loken02, Borgani04, Kay04,
Piffaretti08}).

\begin{table*}
\begin{minipage}[htbp]{\textwidth}
\caption{Parameters of the single \textit{beta}-model, estimated
using the MCMC procedure described in the text, as best fit of the
CC cluster maps.}\label{table_neISO} \centering
\renewcommand{\footnoterule}{}  
  \begin{tabular}{llccccc}
    \hline
     Cluster & FWHM fov & $\Delta T_{SZ0}$ & $\beta$ & $\theta_c$  \\
     & (arcmin) & ($\mu$K) &   & (arcsec)  \\
    \hline
    \hline
   $Cl_{A1413}$ & 1& $-273\pm9$ & $0.731\pm0.006$ &  $66.1\pm3.0$   \\
    & 4.5& $-224\pm14$ & $0.743\pm0.009$ &  $81.7\pm7.3$   \\
    & 7& $-195\pm11$ & $0.758\pm0.011$ &  $96.4\pm8.0$   \\
    \hline
    $Cl_{A1689}$& 1& $-554\pm8$ & $0.979\pm0.006$ &  $80.2\pm1.5$   \\
    & 4.5& $-430\pm12$ & $0.994\pm0.004$ &  $93.7\pm2.2$   \\
    & 7& $-592\pm84$ & $0.909\pm0.018$ &  $61.4\pm8.3$   \\
    \hline
    $Cl_{A1835}$& 1& $-709\pm10$ & $0.951\pm0.006$ &  $56.8\pm1.1$   \\
    & 4.5& $-468\pm29$ & $0.980\pm0.013$ &  $76.0\pm5.0$   \\
    & 7& $-818\pm88$ & $0.870\pm0.010$ &  $36.0\pm3.6$   \\
    \hline
    $Cl_{A2204}$& 1& $-727\pm8$ & $0.848\pm0.003$ &  $68.4\pm0.9$   \\
    & 4.5& $-551\pm18$ & $0.865\pm0.007$ &  $86.0\pm3.6$   \\
    & 7& $-514\pm39$ & $0.859\pm0.017$ &  $86.9\pm9.0$   \\
    \hline
    $Cl_{A2261}$ & 1& $-399\pm10$ & $0.830\pm0.006$ &  $52.3\pm1.7$   \\
    & 4.5& $-273\pm17$ & $0.849\pm0.008$ &  $71.6\pm4.6$   \\
    & 7& $-323\pm32$ & $0.816\pm0.013$ &  $55.5\pm6.4$   \\
    \hline
    $Cl_{MS1358.4+6245}$& 1& $-304\pm12$ & $0.857\pm0.011$ &  $49.6\pm2.7$   \\
    & 4.5& $-245\pm28$ & $0.856\pm0.013$ &  $56.0\pm6.1$   \\
    & 7& $-1246\pm116$ & $0.753\pm0.009$ &  $8.7\pm1.2$   \\
    \hline
    $Cl_{RXJ1347.5-1145}$& 1& $-1683\pm15$ & $0.835\pm0.002$ &  $37.7\pm0.5$   \\
    & 4.5& $-893\pm30$ & $0.860\pm0.005$ &  $62.7\pm2.5$   \\
    & 7& $-739\pm41$ & $0.873\pm0.011$ &  $73.1\pm5.1$   \\
    \hline
    $Cl_{ZW3146}$& 1& $-645\pm13$ & $0.839\pm0.005$ &  $42.4\pm1.1$   \\
    & 4.5& $-404\pm38$ & $0.857\pm0.013$ &  $61.3\pm6.7$   \\
    & 7 & $-585\pm77$ & $0.801\pm0.011$ &  $36.5\pm5.2$   \\
    \hline
  \end{tabular}
\end{minipage}
\end{table*}

The profile, proposed in this paper specifically for CC clusters,
follows both the central drop and the outer decline of the gas
temperature. The function is formalized in the
$\log\left(r/r_{500}\right)-\log\left(T_e(r)/T_X\right)$ plane on
which the power laws of Eq. \ref{eq_Te_general} become linear
functions (see Appendix \ref{sec_appendix} for a complete
treatment). To fix the parameters $A_1$ and $A_2$ of the radial
electron temperature profile and to confirm the power laws indices
$m_1$ and $m_2$, we fit a co-adding of Chandra electron temperature
normalized to $T_X$ data, of the CC clusters selection with the
proposed function. $T_X$ represents the average temperature value in
the range $(0.1\div1.0)r_{500}$, and it is used to scale each
cluster, in order to fit the universal temperature function to the
measured profiles. In Table \ref{table_te} we report the cluster
redshift, $z$, and the angular diameter distance, $D_A$. The scale
radius $r_{500}$ and temperature $T_X$ are also collected. The
chosen radial range, which is used to calculate the scale
temperature $T_X$, corresponds to a cut in the central region
($r<0.1r_{500}$). Obviously this value cannot be compared easily
with results of other works because it strictly depends on the data
radial extension. In fact, an important source of bias is the
temperature definition (\cite{Vikhlinin06}). Here, we use the
spectroscopic temperature $T_X$. In Figure \ref{fig_fitTe} the
temperature data of our cluster sample with the best fit are
plotted. By following Appendix \ref{sec_appendix}, the resulting
$T_e$ profile parameters are $A_1=2.41\pm0.14$, $A_2=0.55\pm0.10$,
$m_1=0.38\pm0.02$ and $m_2=-0.29\pm0.11$, which univocally define
the $T_e(r)$ function. We note that, even if the power law that
describes the outskirts of the cluster temperature distribution
suffers larger uncertainties, $m_1$ and $m_2$ are both compatible,
within one standard deviation, with estimates available in the
literature. For example, Zhang et al. (2008) find $m_1=0.38\pm0.04$,
for $r<0.2r_{500}$, in agreement with Sanderson et al. (2006), who
proposes $m_1=0.4$, for $r<0.1r_{500}$. For radii larger then
$0.2r_{500}$, Zhang et al. (2008) fitted a selected sample of data
from XMM-Newton finding structurally similar behavior to ours with
$m_2=-0.28\pm0.19$.

\section{Pipeline of cluster simulation}\label{sec_MCMC}

The analysis reported in this paper can be summarized in the
following steps:
\begin{itemize}
    \item construction of an SZ signal distortion profile $\Delta T_{SZ}
    (\theta)$, using the ICM information coming from existing X-ray
    observations;
    \item convolution of the cluster SZ map with several instrumental beam profiles;
    \item extraction of the ICM parameters as in the assumptions of the isothermal \textit{beta}-model;
    \item estimation of the cluster total mass $M_{tot}$;
    \item calculation of the bias on cluster total mass due to
    the incorrect description of the ICM.
\end{itemize}

In the first step, we generate angular profiles of the SZ signal
distortion $\Delta T_{SZ} (\theta)$, assuming an observing frequency
of 150 GHz. The electron number density $n_e$ and temperature $T_e$
profiles are constructed using the equations presented in Sect.
\ref{sec_ne_Te}, which we assume to be a good representation of a
cool core ICM. The angular profile of the comptonization parameter
is then evaluated by projecting the electron pressure profile on the
plane orthogonal to the cluster line of sight. The SZ signal is
obtained using Eq. \ref{eq_SZ}.

The $\Delta T_{SZ} (\theta)$ profiles are then convolved by
considering three different instrumental beam profiles modeled as a
Gaussian, corresponding to large single dishes (SPT or ACT) with
$FWHM=1$ arcmin, medium size telescopes (MITO or OLIMPO) with
$FWHM=4.5$ arcmin, and small apertures (Planck) with $FWHM=7$
arcmin.

The errors associated to the convolved $\Delta T_{SZ}(\theta)$
profiles are treated as only due to instrumental noise. An
optimistic choice of the sensitivity, for all the observatories, is
6 $\mu$K/beam, corresponding to the Planck channel at 143 GHz
(\cite{Planck11c}) assuming the necessary integration time on source
for the other experiments. Contaminants are not included in the
study since we wish to assess our ability to extract the mass of the
clusters under ideal conditions.

To simulate the missing knowledge of X-ray observational results, we
ignore cluster morphology and assume the most general model for it:
an isothermal \textit{beta}-model, that expressed in temperature is

\begin{equation}\label{eq_DeltaT_beta}
\Delta T_{SZ}=\Delta
T_{SZ0}{\left(1+\frac{\theta^2}{\theta_c^2}\right)}^{\frac{1}{2}-\frac{3}{2}\beta}.
\end{equation}

We apply the Metropolis Hastings (M-H) algorithm Monte Carlo Markov
Chain (MCMC) procedure to fit this equation (after a convolution
with the corresponding instrumental beam) on the simulated profiles
to extract the parameters $\Delta T_{SZ0}$, $\beta$ and $\theta_c$.
For each cluster, at a fixed field of view (\textit{fov}), we
analyze the accepted set of parameters derived by the MCMC
procedure. The degeneracy among the extracted \emph{beta}-model
parameters affects their uncertainties.

\begin{figure}
  \includegraphics[width=\columnwidth]{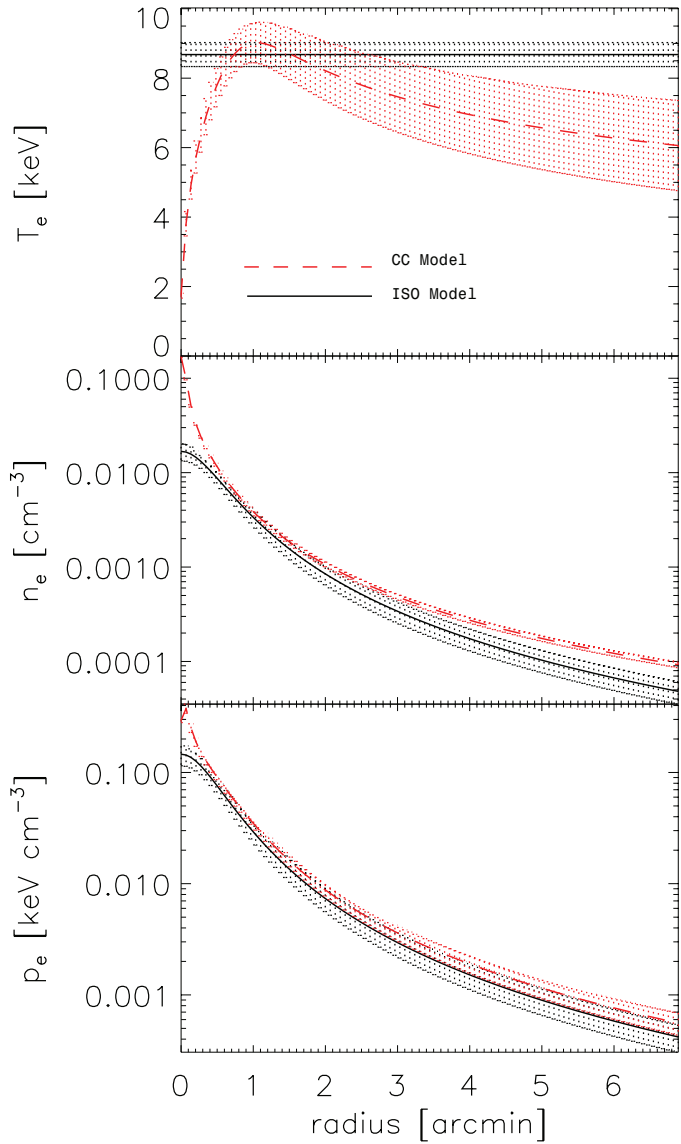}\\
  \caption{\itshape Radial profiles of the electron temperature, $T_e$ (top),
  number density, $n_e$ (middle) and  pressure, $P_e$ (bottom) for
  the cluster ZW3146. The red dashed curve describes the CC template
  while the black solid curve represents the ISO model (the shadowed regions define
  1$\sigma$ uncertainties), with parameters extracted by MCMC analysis
  considering $\Delta T_{SZ}(\theta)$ profiles convolved with a beam
  of 7 arcmin (FWHM).}\label{fig_ne_Te_pe_profiles1}
\end{figure}

We obtain the \textit{beta}-model parameter set that is most
consistent with the $\Delta T_{SZ}(\theta)$ profile, given the
assumed instrumental noise and beam sizes. All the parameters
resulting from the MCMC analysis are collected in Table
\ref{table_neISO} for each cluster and for each \textit{fov}. Figure
\ref{fig_ne_Te_pe_profiles1} shows the electron temperature (top),
number density (middle), and pressure (bottom) profiles for only the
cluster ZW3146, as an example, of both the original CC ICM and the
recovered ISO model. The errors associated to the curves account for
the 1$\sigma$ uncertainties on the parameters.

\section{Cluster total mass estimation}\label{sec_total_mass}

We want to stress the consequences of the assumptions on the ICM
physics when we miss X-ray information. A quantity, such as the
total mass $M_{tot}$, can be biased by a different physical state of
the ICM (i.e. mergers or cooling flow mechanisms). In particular we
estimate the mass for both the ICM discussed templates ($CC$ and
$ISO$), by using the following different approaches:

\begin{itemize}
    \item hydrostatic equilibrium assumption for the cluster
    gas (hydrostatic equilibrium, HE);
    \item gas fraction independence of cluster physical state (fixed gas fraction, FGF);
    \item $M_{tot}-Y$ scaling relation (scaling law,
    SL), as derived in the standard self-similar collapse scenario.
\end{itemize}

The masses are calculated, in particular, within a fixed integration
radius $r_{int}$ (aperture radius), which we arbitrarily choose
equal to the $r_{500}$ values as reported in Morandi et al. (2007,
see Table \ref{table_te}). This choice is motivated by the need to
fix an aperture radius within which to estimate integrated
quantities. We point out that $r_{int}$ does not always correspond
to the same overdensity, due to the different assumed ICM templates.
It is clear that, hereafter, results associated to the clusters
simulated in this work cannot be considered as describing the true
ICM physics of the observed objects. We choose, however, to maintain
the link with the ``native'' cluster in the name ($NAME \rightarrow
Cl_{NAME}$).

\subsection{Hydrostatic equilibrium}

\begin{table*}
\begin{minipage}[htbp]{\textwidth}
\caption{Total cluster masses calculated considering the HE and FGF
approaches.} \label{table_mass HE_FGF} \centering
\renewcommand{\footnoterule}{}  
  \begin{tabular}{llcc}
    \hline
    Cluster & ICM template & $M_{tot,HE}$ & $M_{tot,FGF}$\footnote{derived by $M_{gas}$ assuming
$f_{gas}=0.1$}  \\
     & (FWHM fov)  & (10$^{14}$ $M_{\odot}$)& (10$^{14}$ $M_{\odot}$)  \\
    \hline
    \hline
    A1413 & CC   &$3.85\pm0.77$& $8.73\pm2.03$\\
    $Cl_{A1413}$& ISO (1')   &$6.59\pm0.60$&$6.21\pm0.35$\\
    & ISO (4.5')   &$6.67\pm0.59$&$6.02\pm0.59$\\
    & ISO (7')   &$6.66\pm0.58$&$5.93\pm0.89$\\
    \hline
    A1689 & CC  & $8.96\pm1.97$&$12.72\pm2.33$\\
    $Cl_{A1689}$& ISO (1')  &$13.56\pm1.41$& $9.88\pm3.12$\\
    & ISO (4.5')   &$13.61\pm1.33$&$9.14\pm4.08$\\
    & ISO (7')   &$12.66\pm1.33$&$8.86\pm2.01$\\
    \hline
    A1835 & CC  & $10.23\pm2.30$&$12.72\pm1.18$\\
    $Cl_{A1835}$& ISO (1') &$16.21\pm1.61$& $10.08\pm0.33$\\
    & ISO (4.5')   &$16.32\pm1.55$&$9.13\pm0.94$\\
    & ISO (7')   &$15.08\pm1.40$&$7.99\pm1.44$\\
    \hline
    A2204 & CC  & $12.91\pm3.30$&$14.84\pm2.96$\\
    $Cl_{A2204}$& ISO (1') &$19.88\pm3.27$&$10.13\pm0.24$\\
    & ISO (4.5')   &$20.33\pm3.05$&$10.87\pm0.67$\\
    & ISO (7')   &$19.89\pm2.86$&$10.32\pm1.54$\\
    \hline
    A2261 & CC  & $4.79\pm1.11$&$10.55\pm3.48$\\
    $Cl_{A2261}$& ISO (1')  &$7.88\pm1.09$&$7.80\pm3.50$\\
     & ISO (4.5')   &$7.84\pm1.18$&$7.24\pm1.22$\\
    & ISO (7')   &$7.81\pm1.08$&$6.94\pm1.09$\\
   \hline
    MS1358.4+6245 & CC  & $8.09\pm1.75$&$10.23\pm1.65$\\
    $Cl_{MS1358.4+6245}$& ISO (1')   &$13.27\pm1.77$&$7.74\pm0.63$\\
    & ISO (4.5')   &$13.17\pm1.88$&$7.27\pm1.22$\\
    & ISO (7')   &$11.90\pm1.51$&$6.12\pm1.14$\\
    \hline
    RXJ1347.5-1145 & CC  & $14.66\pm3.02$&$32.34\pm4.05$\\
    $Cl_{RXJ1347.5-1145}$& ISO (1')  &$24.45\pm2.49$& $25.04\pm0.48$\\
    & ISO (4.5')   &$24.47\pm2.54$&$22.08\pm1.30$\\
    & ISO (7')   &$24.66\pm2.49$&$21.07\pm1.84$\\
    \hline
    ZW3146 & CC  & $9.06\pm1.64$&$17.88\pm1.05$\\
    $Cl_{ZW3146}$& ISO (1')  &$15.05\pm0.60$& $13.77\pm0.58$\\
    & ISO (4.5')   &$15.14\pm0.67$&$12.72\pm1.97$\\
    & ISO (7')   &$14.44\pm0.64$&$12.03\pm2.61$\\
    \hline
  \end{tabular}
\end{minipage}
\end{table*}

The first approach, HE, assumes a spherical symmetry for the
cluster, so that the hydrostatic equilibrium equation can be written
(\cite{Sarazin88}) as

\begin{equation}\label{eq_hyd_eq}
\frac{dP_{gas}(r)}{dr}=-\rho_{gas}(r)G\frac{M_{tot}(<r)}{r^2}
\end{equation}
where $M_{tot}(<r)$ is the total cluster mass within radius $r$ and
under the ideal gas assumption, $P_{gas}=(\rho_{gas}k_BT_{gas})/(\mu
m_p)$, with $\rho_{gas}=\mu_e m_p n_e$, $\mu$ and $\mu_e$ are the
total and electron mean molecular weights (i.e. the mean particle
mass per electron in units of the proton mass $m_p$), $G$ is the
gravitational constant and $T_{gas}=T_i=T_e$, the ion and electron
temperatures respectively, because the system is assumed to be in
thermal equilibrium. To adapt this equation to the SZ physics, we
substitute the gas pressure, which accounts for both electrons and
ions, with the simple electron pressure $P_e$, which causes the SZ
effect, so we have

\begin{equation}\label{eq_hyd_eq}
\frac{dP_e(r)}{dr}=-\mu m_p G\frac{M_{tot}(<r)}{r^2}n_e(r).
\end{equation}
The total mass can be derived as

\begin{equation}\label{eq_hyd_eq_Mtot}
M_{tot}(<r)=-\frac{kT_e(r)}{G\mu
m_p}r\left[\frac{\partial\ln(n_e(r))}{\partial\ln(r)}+\frac{\partial\ln(kT_e(r))}{\partial\ln(r)}\right],
\end{equation}
which reduces, in the simple case of the isothermal
\textit{beta}-model, to

\begin{equation}\label{eq_hyd_eq_Mtot_iso}
M_{tot}(<r)=\frac{3\beta kT_e}{G\mu m_p}\frac{r^3}{r^2_c+r^2}.
\end{equation}
The previous equations yield $M_{tot,HE,CC}$ and $M_{tot,HE,ISO}$,
respectively. If we also calculate the cluster gas mass by

\begin{equation}\label{eq_Mgas}
M_{gas}=\mu_em_p\int n_edV,
\end{equation}
\begin{figure}
 \includegraphics[width=\columnwidth]{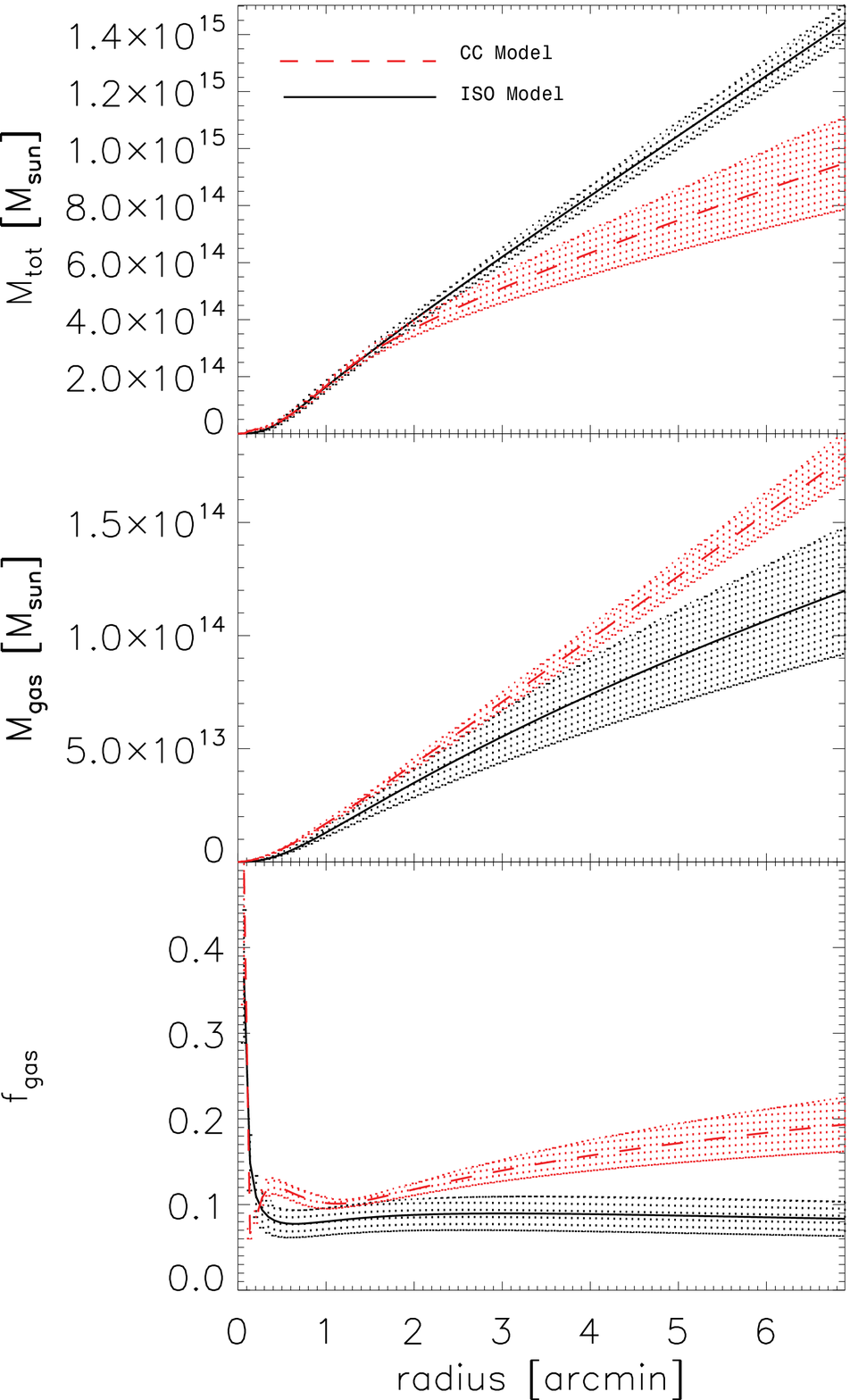}\\
  \caption{\itshape Radial profiles of the cumulative total (top) and
  gas (middle) cluster masses and the gas fraction (bottom) for the cluster $Cl_{ZW3146}$. The
  plots refer to the two ICM templates (CC and ISO) discussed
  in the text, under the hydrostatic hypothesis. The ISO model parameters have been extracted by MCMC analysis
  considering $\Delta T_{SZ}(\theta)$ profiles convolved with a beam
  of 7 arcmin (FWHM).}\label{fig_Mtot_Mgas_log}
\end{figure}
it is easy to estimate the gas fractions, $f_{gas,HE,CC}$, and
$f_{gas,HE,ISO}$. The plots in Fig. \ref{fig_Mtot_Mgas_log} refer,
still as an example, to the cluster $Cl_{ZW3146}$ and show the
radial profiles of the cumulative cluster total and gas masses for
both the ICM templates and the gas fraction. The $1\sigma$ lines are
derived by considering the uncertainties of the ICM parameters.

\subsection{Fixed gas fraction}

The second approach, FGF, assumes that the gas fraction, estimated
at $r_{int}$, is independent of the cluster dynamical state, as
deduced by simulations (e.g. \cite{Rasia06, Lau09}) and as recently
derived by XMM-Newton and Subaru observations (\cite{Zhang10}).
Under this hypothesis, we can derive the cluster total masses
directly by gas masses as $M_{tot,FGF}=M_{gas}/f_{gas}$.

Because the integrated quantities are calculated by referring to a
fixed radius $r_{int}$, defined in the previous section, it is
evident that, for each cluster and for each considered ICM model, we
are dealing with different overdensities. We calculate these
overdensities by

\begin{equation}\label{eq_delta}
\Delta_{int}=\frac{2G}{H_0^2E(z)^2r_{int}^3}M_{tot}(r_{int}).
\end{equation}

The results obtained for the HE and FGF approaches are collected in
Table \ref{table_mass HE_FGF}. In order to derive all the quantities
corresponding to the ISO template from the parameters obtained with
the MCMC procedure, we need to convert the $\Delta T_{SZ0}$ values
to central electron number densities. This is possible using
equation

\begin{equation}\label{eq_n_e0}
n_{e0}=\frac{\Delta T_{SZ0}}{T_{CMB}f(x)}\frac{m_ec^2}{\sigma_T
\theta_cD_A\sqrt{\pi}}\frac{\Gamma(\frac{3}{2}\beta)}{\Gamma(\frac{3}{2}\beta-\frac{1}{2})}\frac{1}{T_{e0}},
\end{equation}
which is derived by expressing Eq. \ref{eq_y0} in terms of $\Delta
T_{SZ}$. Of course this expression introduces a degeneracy between
temperature and density, possibly producing the same $\Delta
T_{SZ0}$ value. For this reason the masses presented in this work
have been obtained by fixing the electron temperature to the $T_X$
values (Table \ref{table_te}) and assuming $f_{gas}=0.1$.

\subsection{$M_{tot}-Y$ scaling relation}

\begin{table*}
\begin{minipage}[htbp]{\textwidth}
\caption{Total mass as derived by applying the $M_{tot}-Y$ scaling
law as in Eq. \ref{eq_scaling_M-Ycyl}.} \label{table_SL} \centering
\renewcommand{\footnoterule}{}  
  \begin{tabular}{llcccc}
    \hline
    Cluster &ICM template& $D_A^2Y$ &$f_{gas}$ & $\Delta_{int}$& $M_{tot,SL}$ \\
    name & (FWHM fov)& (kpc$^2$) & & & (10$^{14}$ $M_{\odot}$) \\
    \hline
    \hline
    A1413 & CC& $100\pm1$ & $0.23\pm0.05$ & $345\pm69$& $1.94\pm0.23$ \\
    $Cl_{A1413}$& ISO (1')& $95\pm5$ & $0.10\pm0.01$ & $591\pm53$& $2.81\pm0.18$ \\
     & ISO (4.5')& $92\pm12$ & $0.09\pm0.01$ & $598\pm53$& $2.86\pm0.30$ \\
     & ISO (7')& $90\pm9$ & $0.09\pm0.01$ & $597\pm52$& $2.81\pm0.26$ \\
    \hline
    A1689 & CC& $177\pm2$ & $0.15\pm0.03$ & $477\pm105$& $3.34\pm0.38$ \\
     $Cl_{A1689}$&ISO (1')& $152\pm6$ & $0.07\pm0.01$ & $722\pm75$& $4.15\pm0.23$ \\
     &ISO (4.5')& $142\pm6$ & $0.07\pm0.01$ & $724\pm71$& $4.17\pm0.23$ \\
     &ISO (7')& $139\pm32$ & $0.07\pm0.02$ & $674\pm71$& $4.16\pm0.82$ \\
    \hline
    A1835 & CC& $263\pm2$ & $0.13\pm0.03$ & $468\pm105$& $4.48\pm0.36$ \\
     $Cl_{A1835}$&ISO (1')& $186\pm7$ & $0.06\pm0.01$ & $741\pm74$& $5.05\pm0.25$ \\
      &ISO (4.5')& $171\pm17$ & $0.06\pm0.01$ & $746\pm71$& $5.12\pm0.51$ \\
      &ISO (7')& $152\pm28$ & $0.05\pm0.01$ & $689\pm64$& $5.00\pm0.85$ \\
    \hline
     A2204 & CC& $271\pm2$ & $0.12\pm0.04$ & $337\pm86$& $5.28\pm0.72$ \\
     $Cl_{A2204}$&ISO (1')& $241\pm6$ & $0.06\pm0.01$ & $520\pm86$& $6.78\pm0.47$ \\
     &ISO (4.5')& $231\pm15$ & $0.05\pm0.01$ & $531\pm80$& $6.89\pm0.59$ \\
      &ISO (7')& $222\pm31$ & $0.05\pm0.01$ & $520\pm75$& $6.88\pm0.99$ \\
    \hline
    A2261 & CC& $134\pm2$ & $0.22\pm0.06$ & $388\pm90$& $2.31\pm0.38$ \\
     $Cl_{A2261}$&ISO (1')& $110\pm5$ & $0.10\pm0.02$ & $639\pm89$& $2.87\pm0.19$ \\
    &ISO (4.5')& $104\pm9$ & $0.09\pm0.02$ & $635\pm95$& $2.93\pm0.29$ \\
     &ISO (7')& $99\pm16$ & $0.09\pm0.02$ & $633\pm88$& $2.93\pm0.48$ \\
    \hline
    MS1358.4+6245& CC& $166\pm4$ & $0.13\pm0.03$ & $233\pm50$& $3.82\pm0.35$ \\
   $Cl_{MS1358.4+6245}$&ISO (1')& $127\pm11$ & $0.06\pm0.01$ & $381\pm51$& $4.67\pm0.45$ \\
   &ISO (4.5')& $113\pm21$ & $0.06\pm0.01$ & $379\pm54$& $4.56\pm0.76$ \\
   &ISO (7')& $96\pm18$ & $0.05\pm0.01$ & $342\pm43$& $4.36\pm0.76$ \\
     \hline
   RXJ1347.5-1145 & CC& $1165\pm4$ & $0.23\pm0.03$ & $305\pm63$& $8.09\pm0.52$ \\
      $Cl_{RXJ1347.5-1145}$&ISO (1')& $719\pm14$ & $0.10\pm0.01$ & $509\pm52$& $8.68\pm0.40$ \\
      &ISO (4.5')& $654\pm41$ & $0.09\pm0.01$ & $509\pm52$& $8.86\pm0.60$ \\
      &ISO (7')& $649\pm66$ & $0.09\pm0.01$ & $513\pm52$& $9.11\pm0.80$ \\
     \hline
   ZW3146 & CC& $312\pm4$ & $0.20\pm0.04$ & $201\pm37$& $4.42\pm0.31$ \\
     $Cl_{ZW3146}$&ISO (1')& $224\pm11$ & $0.09\pm0.01$ & $334\pm13$& $5.21\pm0.23$ \\
     &ISO (4.5')& $208\pm38$ & $0.08\pm0.01$ & $336\pm15$& $5.31\pm0.81$ \\
     &ISO (7')& $205\pm43$& $0.08\pm0.02$ & $321\pm14$ & $5.41\pm0.11$ \\
    \hline
  \end{tabular}
\end{minipage}
\end{table*}

The total cluster mass can be inferred by applying a self-similar
relation that links it to the integrated comptonization parameter.
The $M_{tot}-Y$ scaling law is usually calibrated for a fixed
overdensity value, while in this work we are dealing with masses
calculated at different overdensities. For this reason in the
scaling law, we make the dependence on overdensity explicit. Simple
considerations, based on the assumption that cluster evolution is
completely determined by gravitational processes, lead to an easy
scaling relation that connects the total mass of a cluster of
galaxies, $M_{tot}$, to its temperature, $T_e$, considering an
isothermal structure for the ICM. Following Kravtsov et al. (2006)
and Bryan \& Norman (1998) and keeping the overdensity dependence
explicit, we have

\begin{equation}\label{eq_scaling_M-T}
k_BT_e=\mu m_p{\left[\frac{27}{16}\Delta_{int}
G^2H_0^2E(z)^2\right]}^{1/3}M_{tot}^{2/3}.
\end{equation}

To connect a spherically integrated quantity (e.g. cluster total
mass) to one derived by a cylindrical integration (e.g. integrated
SZ flux), we consider the spherical analogous to the SZ flux,
$Y_{S}$. This quantity is directly proportional to the cluster total
mass as

\begin{equation}\label{eq_scaling_M-Y-T}
Y_{S}=\frac{k_B\sigma_T}{m_ec^2}\int
n_eT_edV=\frac{k_B\sigma_T}{m_ec^2}T_{mw}
\frac{f_{gas}M_{tot}}{\mu_e m_p}
\end{equation}
where $M_{tot}$ is the cluster total mass inside a sphere with
radius equal to $r_{int}$, and $T_{mw}$ is the gas mass-weighted
mean temperature.

By combining the previous equations, we get

\begin{equation}\label{eq_scaling_M-Ysph}
M_{tot}=0.248{\left[f_{gas}^3E(z)^2\Delta_{int}\right]}^{-1/5}Y_{S}^{3/5}\mbox{$10^{14}$
$M_{\odot}$}.
\end{equation}
This means that, by estimating $Y_{S}$, we can easily infer the
corresponding  total cluster mass. Furthermore, in order to derive
the integrated SZ flux, we need to solve the equation

\begin{equation}
Y=2\pi\int^{\theta_{int}}_0 y(\theta)\theta d\theta,
\end{equation}
where $\theta_{int}=r_{int}/D_A$ and $y(\theta)$ is extracted by the
simulated SZ temperature decrement profiles (in the CC case) or
calculated directly from Eq. \ref{eq_ybeta}, by considering the
parameters as derived from the MCMC analysis in Sect. \ref{sec_MCMC}
(in the ISO case). The two defined integrated Comptonization
parameters ($Y_S$ and $Y$) are connected by the dimensionless
quantity $C=D_A^2Y/Y_{S}$.

We averaged the $C$ factors, assuming different cylindric depths
equal to 2, 5, and 10 $r_{int}$, obtaining $C_2=1.52\pm0.07$,
$C_5=1.85\pm0.14$, and $C_{10}=2.00\pm0.19$, respectively,
converging to the value in Bonamente et al. (2008).

The scaling law can be rewritten as

\begin{equation}\label{eq_scaling_M-Ycyl}
M_{tot}=0.248{\left[f_{gas}^3E(z)^2
C_{\infty}^3\Delta_{int}\right]}^{-1/5}(D_A^2Y)^{3/5}\mbox{$10^{14}$
$M_{\odot}$},
\end{equation}
which can directly give an estimation of cluster total mass
$M_{tot,SL}$. All the quantities included in the scaling law are
listed in Table \ref{table_SL}

\section{Bias on the total mass}\label{sec_MB}

\begin{table*}
\begin{minipage}[htbp]{\textwidth}
\caption{Biases on the cluster total mass estimate for three
different approaches as derived by three different fovs.}
\label{table_bias} \centering
\renewcommand{\footnoterule}{}  
  \begin{tabular}{l l c c c}
    \hline
    Cluster& FWHM fov & $MB_{HE}$ & $MB_{FGF}$ & $MB_{SL}$ \\
    name& (arcmin) & & &\\
    \hline
    \hline
    $Cl_{A1413}$& 1 & $0.75\pm0.31$ & $-0.29\pm0.19$ & $0.58\pm0.32$\\
    & 4.5 & $0.79\pm0.35$ & $-0.30\pm0.15$ & $0.59\pm0.34$\\
    & 7 & $0.81\pm0.31$ & $-0.33\pm0.17$ & $0.61\pm0.31$\\
    \hline
    $Cl_{A1689}$& 1 & $0.56\pm0.31$ & $-0.24\pm0.14$ & $0.43\pm0.24$\\
    & 4.5 & $0.59\pm0.33$ & $-0.30\pm0.15$ & $0.45\pm0.28$\\
    & 7 & $0.50\pm0.30$ & $-0.35\pm0.20$ & $0.55\pm0.45$\\
    \hline
    $Cl_{A1835}$& 1 & $0.63\pm0.31$ & $-0.26\pm0.07$ & $0.44\pm0.16$\\
    & 4.5 & $0.64\pm0.28$ & $-0.32\pm0.09$ & $0.47\pm0.21$\\
    & 7 & $0.51\pm0.32$ & $-0.41\pm0.11$ & $0.45\pm0.29$\\
    \hline
    $Cl_{A2204}$& 1 & $0.59\pm0.32$ & $-0.25\pm0.15$ & $0.45\pm0.27$\\
    & 4.5 & $0.62\pm0.30$ & $-0.29\pm0.15$ & $0.50\pm0.27$\\
    & 7 & $0.63\pm0.30$ & $-0.31\pm0.18$ & $0.47\pm0.33$\\
    \hline
    $Cl_{A2261}$&1  & $0.75\pm0.30$ & $-0.23\pm0.25$ & $0.52\pm0.36$\\
    & 4.5 & $0.74\pm0.33$ & $-0.28\pm0.29$ & $0.53\pm0.40$\\
    & 7 & $0.69\pm0.32$ & $-0.30\pm0.25$ & $0.54\pm0.40$\\
    \hline
    $Cl_{MS1358.4+6245}$& 1 & $0.71\pm0.31$ & $-0.28\pm0.12$ & $0.50\pm0.21$\\
    & 4.5 & $0.69\pm0.34$ & $-0.32\pm0.17$ & $0.55\pm0.34$\\
    & 7 & $0.52\pm0.29$ & $-0.43\pm0.14$ & $0.47\pm0.32$\\
    \hline
    $Cl_{RXJ1347.5-1145}$& 1 & $0.79\pm0.32$ & $-0.26\pm0.08$ & $0.51\pm0.17$\\
    & 4.5 & $0.77\pm0.30$ & $-0.35\pm0.08$ & $0.56\pm0.19$\\
    & 7 & $0.82\pm0.38$ & $-0.38\pm0.10$ & $0.61\pm0.26$\\
    \hline
    $Cl_{ZW3146}$& 1 & $0.66\pm0.32$ & $-0.28\pm0.06$ & $0.47\pm0.16$\\
    & 4.5 & $0.67\pm0.32$ & $-0.33\pm0.12$ & $0.53\pm0.27$\\
    & 7 & $0.62\pm0.33$ & $-0.40\pm0.14$ & $0.57\pm0.31$\\
    \hline
  \end{tabular}
\end{minipage}
\end{table*}

The results clearly show that a general interpretation of cluster
physics, as assumed in a pure self-similar scaling law like
$M_{tot}-Y$, can infer a wrong estimate of cluster total mass. To
emphasize  and to quantify this point, we define a mass bias as

\begin{equation}\label{eq_MB}
MB=\frac{M_{tot,ISO}-M_{tot,CC}}{M_{tot,CC}}.
\end{equation}
This quantifies the difference between the SZ derived mass, as
results from the isothermal \textit{beta}-model assumption (ISO) and
the X-ray derived mass (CC), which we deduce from the HE, FGF, and
SL approaches. We notice no significant bias dependence on the beam
size used to convolve the SZ signals, under the assumption of the
presence of instrumental noise alone. Therefore, all plots refers to
results obtained with a \textit{fov} of 7 arcmin FWHM, considering
$T_X$ values as in Table \ref{table_te}. We calculate the mass bias
for all the analyzed approaches.

\begin{figure*}
\centering
  \begin{flushleft}\hspace{2.5cm} \Large{(a)} \hspace{4.6cm} \Large{(b)} \hspace{3.0cm} \Large{(c)}  \hspace{3.0cm} \Large{(d)}\\\end{flushleft}
  \includegraphics[width=\textwidth]{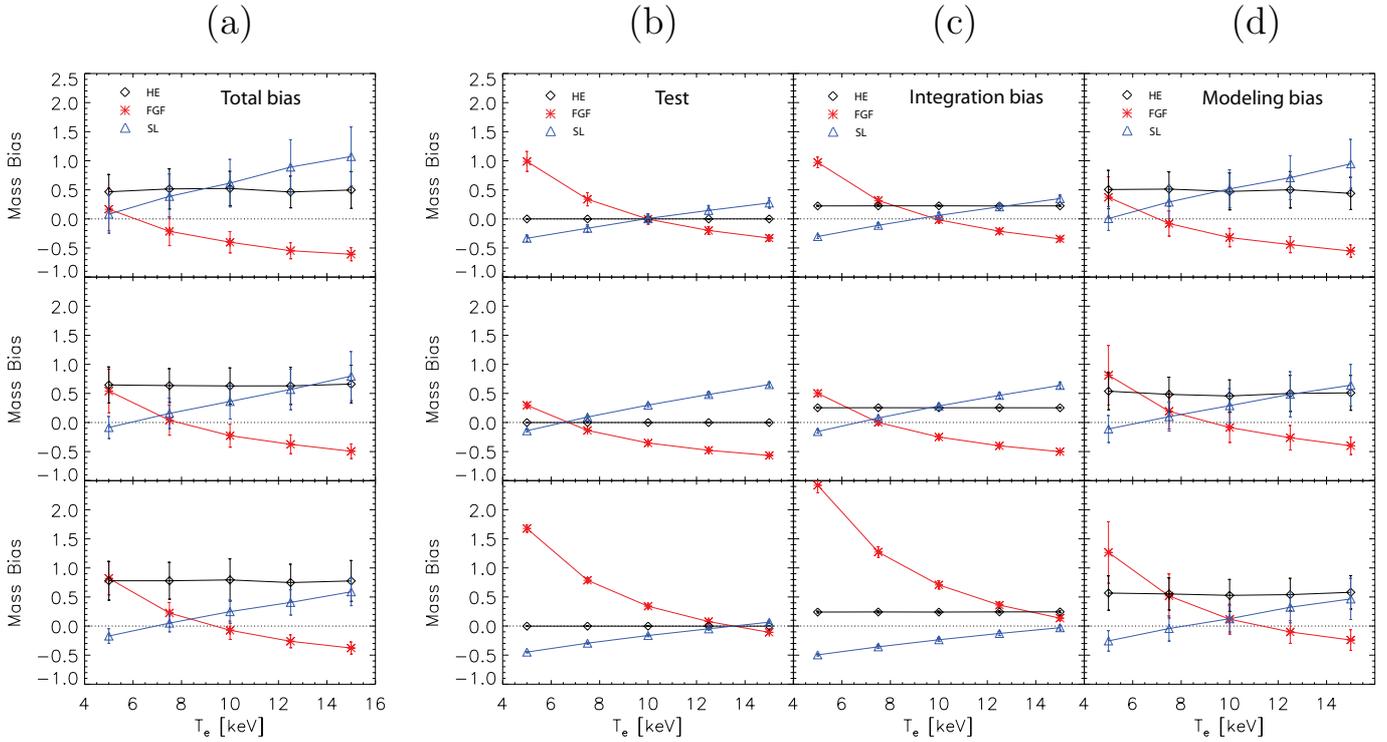}
  \caption{\itshape Total mass biases estimated
  with different approaches, HE (black), FGF (red), SL (blue), for three representative
  clusters of our sample (from top to bottom: $Cl_{A1689}$,
  $Cl_{A2204}$, and
  $Cl_{RXJ1347.5-1145}$). The ISO parameters used for the bias
  estimation are associated to SZ signals convolved with a beam
  of 7 arcmin (FWHM). In panel (a) the total mass bias is shown, while in
  panel (b) the reliability of bias estimation has been tested. The two
  bias contributions are disentangled: integration (c) and modeling (d)
  biases. Assuming no prior for $T_e$, the mass biases are plotted
  versus a reliable range of electron temperatures (5-15 keV).}\label{fig_bias}
\end{figure*}

Since the $\Delta T_{SZ0}$ parameter does not give us unique
information on the physics of the ICM, we have to study different
pairs of the parameters $n_{e0}$ and $T_x$. Thus, we select values
of $T_x$ that describe a reliable range of electron temperatures
(from 5 keV to 15 keV) and calculate the corresponding $n_{e0}$. The
mass biases are plotted in Figure \ref{fig_bias}a, where the three
cases (HE, FGF, and SL) are shown all together in the plots
corresponding to clusters $Cl_{A1689}$, $Cl_{A2204}$, and
$Cl_{RXJ1347.5-1145}$, representing a wide span in the electron
temperature values. Table \ref{table_bias} lists the mass biases, as
estimated at the X-ray derived electron temperature ($T_X$ in Table
\ref{table_te}).

A check of the goodness of the procedure was done by simulating the
observation of a cluster having an isothermal \textit{beta}-model
$\Delta T_{SZ}(\theta)$ profile instead of a CC cluster. For this
validation procedure, named \textit{Test}, we used the same
analytical expression of the SZ signal to extract the parameters.
The assumed electron temperature is the one obtained by Bonamente et
al. (2006) and reported in the last column in Table \ref{table_te}.
In Figure \ref{fig_bias}b we represent the cluster total mass bias
with these assumptions. While the bias is always zero in the HE
approach, for the FGF and SL cases, as expected, we notice a mass
bias dependence on $T_e$. Due to the degeneracy between the electron
temperature and number density, the mass bias varies with $n_e$. In
all cases, it is worth noting that these biases nullify for electron
temperatures equal to $T_{e0}$ values, thus proving that the method
is not affected by systematics.

The degeneracy between the ICM parameters, resulting from yielding
the same SZ signal, produces different trends on FGF and SL biases
with electron temperature. For increasing values of electron
temperature it underestimates the $M_{gas}$. Considering the FGF
approach, this implies an underestimation of the total mass, too.
For the SL method this would instead produce an increasing trend in
the mass bias with temperature because an $M_{gas}$ underestimation
corresponds to a decreasing gas fraction.

The net result is a mass bias that is always different from zero for
the HE approach, as well as for the FGF and SL ones, unless it is
for an electron temperature suitable to correctly modeling the CC
cluster as an isothermal one but generally different from the true
ICM temperature (see the crossovers in mass biases in Figure
\ref{fig_bias}).

The mass bias reflects a combination of different contributions
(\cite{Piffaretti08}). In particular for this work, the bias can be
reduced to an \textit{integration bias} and a \textit{modeling
bias}, which are estimated and discussed separately. The
\textit{integration bias} refers to the mass bias due to the
different integration ranges along the line of sight. An infinite
integration range is considered for the analytic $\Delta T_{SZ}$
expression, while a finite range is assumed in the numerical
projection method, which is limited by the missing knowledge of the
electron pressure profile outside the region defined by the data.
The \textit{integration bias} that can be evaluated by still
applying the whole procedure to an ISO template (Figure
\ref{fig_bias}c). The \textit{modeling bias} reflects the dependence
of the bias on the assumed model for the ICM. In order to highlight
it and to cancel the \textit{integration bias}, we apply the
procedure to the CC cluster, recovered as an isothermal one. In this
case both the cluster signals (CC mock data and ISO recovered data)
are obtained by projecting the three-dimensional electron pressure
profile. For disentangling the \textit{integration bias}, the ISO
profile is also integrated over a limited range along the line of
sight. The HE bias ranges between $50\%$ and $80\%$, always implying
a mass overestimation that is independent of the electron
temperature.

\section{Conclusions}\label{sec_conclusions}

We studied the bias that affects the estimate of cluster total mass
by SZ observations when an isothermal \textit{beta}-model is assumed
to describe the ICM physical properties, specifically in the case of
CC clusters when X-ray and/or lensing information is missing.

It is well known that, rather than the central Comptonization
parameter $y_0$, affected by the choice of cluster profile modeling,
an integrated quantity, like the parameter $Y$, appears to be a more
robust mass proxy. Nevertheless, we have shown that CC clusters can
generate observed $y$ maps in which the ICM morphology could still
be substantially hidden, even for the current most sensitive
experiments operating from the largest available mm/submm
telescopes.

While a general assumption of cluster morphology is efficient at
detecting them in blind SZ maps, the possible mismatch with the
actual cluster profile results in a mass bias. In fact simple ICM
models, like the isothermal \textit{beta}-model, applied to SZ
observations can wrongly estimate cluster total mass in the presence
of peculiar ICM dynamics as in CC clusters, which are studied in the
current analysis, and mergers.

We analyzed the mass bias as derived in a limited sample of eight CC
clusters observed by Chandra, both nearby ($0.1<z<0.5$) and with
high mass ($M>10^{14}$ $M_{\odot}$). Under the assumption of an
isothermal \textit{beta}-model, the cluster total mass was derived
applying three different approaches: the hydrostatic equilibrium
equation, a fixed gas fraction, and a self-similar $M_{tot}-Y$
relation.

Assuming we had no information from X-ray observations, we reported
the bias on the derived total mass as dependent on electron gas
temperature. Only in the case of hydrostatic equilibrium does this
bias appear almost constant for the considered clusters in the range
of 50-80 \%. Incidentally, we notice that an electron temperature
value exists for which the FGF and SL mass biases vanish. This could
be the only case in which a simple isothermal \textit{beta}-model
accurately reproduces the mass of CC clusters.

The large biases on total cluster mass recovery in CC clusters
represent another reason to definitely discard the isothermal
\textit{beta}-model for this purpose and to firmly support more
sophisticated models, with universal pressure profiles (e.g.
\cite{Arnaud10}). This is already employed for modeling cluster
atmospheres in almost all the present blind-survey data reduction
(SPT and Planck), and it is planned in the next future for ACT
observations.

\begin{acknowledgements}
Part of this work has been supported by funding from Ateneo
2009-C26A09FTJ7. We thank S. Borgani and the anonymous referee for
their useful comments and suggestions.
\end{acknowledgements}

\appendix

\section{Electron temperature profile function for cool core clusters}\label{sec_appendix}

Here we describe the simple approach used to obtain the electron
temperature profile function applied in this work. The assumptions
at the basis of this treatment are
\begin{itemize}
    \item the knowledge of functional relations of the electron temperature
    in two different regions of the cluster (see Sect.
    \ref{sec_ne_Te}), and
    \item the approximate location of the maximum of the $T_e$ profile.
\end{itemize}

Equation \ref{eq_Te_general} can be also written as

\begin{eqnarray}\label{eq_Te_log}
\log\left(\frac{T^{CC}_e(r)}{T_x}\right)=\left\{ \begin{array}{lr}
m_1\log\left(\frac{r}{r_{500}}\right)+q_1 & \mbox{  for } \frac{r}{r_{500}}<0.1\\
m_2\log\left(\frac{r}{r_{500}}\right)+q_2 & \mbox{  for }
\frac{r}{r_{500}}>0.2\\
\end{array}\right.
\end{eqnarray}
which is clearly obtained by self-similarity studies of $T_e$
profiles and by assuming a universal $T_e$ profile describing CC
clusters. Thus, we need a function that follows those linear trends
asymptotically. Following Eq. \ref{eq_Te_log} it is simple to
associate $A_{1,2}=10^{q_{1,2}}$. For the sake of simplicity, in
what follows we call $Y=\log\left(T_e(r)/T_X\right)$ and
$X=\log\left(r/r_{500}\right)$.

Though several functions can satisfy these general conditions, we
consider the hyperbole as the simplest candidate for yielding a
continuum function. It is extremely easy to find the hyperbole
equation by knowing its asymptotes, only if they are symmetric with
respect to the coordinated axes. In order to be in this simple case,
we need to change our reference frame into a more convenient one,
which results in translating the old reference frame into the
intersections point of the two linear functions and by rotating it
by an angle $\alpha_R=0.5\left[\arctan(m_1)+\arctan(m_2)\right]$,
which is the angle between the old X-axis and the bisectrix of the
asymptotes (i.e. the new X-axis). The new asymptotes, which are now
referred to the new reference frame, are

\begin{eqnarray}
Y_1^{TR}&=&\left[\frac{m_1\cos\alpha_R-\sin\alpha_R}{\cos\alpha_R+m_1\sin\alpha_R}\right]X=m_{R1}X=m_RX\\
Y_2^{TR}&=&\left[\frac{m_2\cos\alpha_R-\sin\alpha_R}{\cos\alpha_R+m_2\sin\alpha_R}\right]X=m_{R2}X=-m_RX
\end{eqnarray}
where the $TR$ apex indicates that the equations are translated
($T$) and rotated ($R$). Now that we have the two symmetric
asymptotes with respect to the coordinated axes, we can use the
hyperbole equation to build up our function, which reads

\begin{equation}\label{eq_asymptotes_RT}
Y^{TR}=\pm\sqrt{\left(\frac{b}{a}\right)^2X^2+b^2}
\end{equation}
where $a=c/\sqrt{1+m_R^2}$, $b=a m_R$ and $c=\sqrt{a^2+b^2}$. Since
the hyperbole of Eq. \ref{eq_asymptotes_RT} is now related to the
wrong asymptotes, we need to inverse-translate it and inverse-rotate
it. The resulting functions are

\begin{eqnarray}
Y^{TRR^{-1}}&=&\frac{-BBX\pm\sqrt{BB^2X^2-4AA(CCX^2-b^2)}}{2AA}\\
Y^{TRR^{-1}T^{-1}}&=&\frac{-BB(X-X_P)}{2AA}\pm\nonumber\\
&\pm&\frac{\sqrt{BB^2(X-X_P)^2-4AA[CC(X-X_P)^2-b^2]}}{2AA}\nonumber\\
&+&Y_P
\end{eqnarray}
where

\begin{eqnarray}
AA&=&\left[\cos^2(\alpha_R)-\left(\frac{b}{a}\right)^2\sin^2(\alpha_R)\right]\nonumber\\
BB&=&\left[1+\left(\frac{b}{a}\right)^2\right]\sin2\alpha_R\nonumber\\
CC&=&\left[\sin^2(\alpha_R)-\left(\frac{b}{a}\right)^2\cos^2(\alpha_R)\right]\nonumber\\
\end{eqnarray}
and

\begin{eqnarray}
X_P&=&\frac{q_2-q_1}{m_1-m_2}\nonumber\\
Y_P&=&\frac{m_1q_2-m_2q_1}{m_1-m_2}\nonumber.
\end{eqnarray}

The parameters $m_1$, $m_2$, $q_1$, $q_2$, characterize the
asymptotes of the discussed hyperbole. The complete function can be
obtained by substituting $(X,Y)$ with
$\left(\log\left(r/r_{500}\right),\log\left(T_e(r)/T_X\right)\right)$
and only considering the negative sign. We add a fifth degree of
freedom identified with $c$ to the analysis, which gives the focal
point position of the hyperbole, and we arbitrarily consider in the
analysis $c=0.2$, which is the one that gives a minimum $\chi^2$ in
the fit with the cluster electron temperature data.

\end{document}